\documentclass[
    ,final            
  ]
  {aipproc}

\layoutstyle{6x9}

\def\xv{{\bf x}}
\begin{document}

\title{Boundary-induced heterogeneous absorbing states}


\author{Juan A. Bonachela} 
{address={Departamento de
  Electromagnetismo y F{\'\i}sica de la Materia and \\ Instituto de
  F{\'\i}sica Te{\'o}rica y Computacional Carlos I, Facultad de
  Ciencias, Universidad de Granada, 18071 Granada, Spain} }

\author{Miguel A.  Mu\~noz}
{address={Departamento de
  Electromagnetismo y F{\'\i}sica de la Materia and \\ Instituto de
  F{\'\i}sica Te{\'o}rica y Computacional Carlos I, Facultad de
  Ciencias, Universidad de Granada, 18071 Granada, Spain} }

\begin{abstract}
  We study two different types of systems with many absorbing states
  (with and without a conservation law) and scrutinize the effect of
  walls/boundaries (either absorbing or reflecting) into them.  In
  some cases, non-trivial structured absorbing configurations
  (characterized by a background field) develop around the
  wall.  We study such structures using a mean-field approach as well
  as computer simulations. The main results are: i) for systems in the
  directed percolation class, a very fast (exponential) convergence of
  the background to its bulk value is observed; ii) for systems with a
  conservation law, power-law decaying landscapes are induced by both
  types of walls: while for absorbing walls this effect is already
  present in the mean-field approximation, for reflecting walls the
  structured background is a noise-induced effect. The landscapes are
  shown to converge to their asymptotic bulk values with an exponent
  equal to the inverse of the bulk correlation length exponent.
  Finally, the implications of these results in the context of
  self-organizing systems are discussed.
\end{abstract}

\maketitle

\section{Introduction: 
Boundaries in systems with absorbing states}

Systems with absorbing states played a dominant role in the
development of non-equilibrium statistical physics \cite{AS}. Directed
percolation, the contact process, or the Domany-Kinzel automaton,
among many other similar models, have been studied profusely. They all
exhibit a phase transition from an active phase, with indefinitely
sustained non-trivial dynamics, to an absorbing phase in which the
system falls with certainty into a frozen state in which all dynamics
ceases.  Applications run from epidemics, to flow in porous media,
auto-catalytic reactions, self-organization, damage spreading,
population dynamics, turbulence, etc.

The critical behavior of most of these systems yields into the very
robust directed percolation (DP) universality class \cite{conjecture},
described at a coarse-grained level by the Reggeon field theory or
Gribov process \cite{AS}. An experimental realization of DP has been
recently obtained, in a breakthrough work, by Takeuchi et al.
\cite{Takeuchi}.  It is only in the presence of extra symmetries,
long-range interactions, or conservation laws, that critical behavior
different from DP can be observed \cite{AS,Lubeck}.

There are important physical situations in which the number of
absorbing configurations grows exponentially with system size, being
infinite in the thermodynamic limit.  Two main universality classes of
such systems are:
\begin{enumerate}
\item The {\it directed percolation class with many AS} \cite{many}.
  Defined by models as the pair contact process, the threshold
  transfer process, and models of catalytic surface reactions
  \cite{models}, this class has no extra symmetry/conservation-law
  with respect to DP. Its corresponding Langevin equation is:
\begin{eqnarray}
  \displaystyle{\partial_t \rho (\xv,t)}& = &
  a \rho - b \rho^2 + \gamma \rho \Psi(\xv,t) +  \nabla^2 \rho +
  \sigma \sqrt{\rho} \eta(\xv,t) \nonumber \\
  \displaystyle{\partial_t \Psi (\xv,t)}& = &
 \alpha \rho - \beta \Psi \rho + D \nabla^2 \rho 
\label{PCP}
\end{eqnarray}
where $a, b, \gamma, \alpha$ and $\beta$ are constants and $\eta$ is a
Gaussian white noise. Despite the non-trivial absorbing phase,
characterized by the non-diffusive background-field $\Psi(\xv,t)$, it
exhibits DP-like (bulk) criticality (see \cite{many} for more
details).

\item The {\it C-DP} class, introduced to describe the criticality of
  stochastic sandpiles, as the {\it Manna} or the {\it Oslo} one
  \cite{Sandpiles,SOC1,SOC2}, is represented by a Langevin equation with an
  extra conservation law, rendering it different from DP
  \cite{FES,Romu,Lubeck}:
\begin{eqnarray}
  \displaystyle{\partial_t \rho (\xv,t)}& = &
  a \rho - b \rho^2 + \omega \rho E(\xv,t) +  \nabla^2 \rho +
  \sigma \sqrt{\rho} \eta(\xv,t) \nonumber \\
  \displaystyle{ \partial_t E (\xv,t)}& = & D \nabla^2 \rho.
\label{CDP}
\end{eqnarray}
\noindent
$a$, $b$, $D$ and $\omega$ are parameters and $E(\xv,t)$ is the
(conserved and non-diffusive) background, usually called {\it energy
  field}. Higher order, irrelevant, terms have been omitted.
\end{enumerate}

It has been claimed, and confirmed in various ways, that the
prototypical models of self-organized criticality (SOC), i.e. {\it
  sandpiles} \cite{Sandpiles}, fluctuate around a critical point owing
to the combined effect of slow driving (addition of energy) and open
boundaries (energy dissipation), and that such a critical point is in
the C-DP class, owing to the conserved nature of the (bulk)
redistribution dynamics \cite{FES,GG}.

A key issue to fully clarify this connection is to elucidate whether
the heterogeneity introduced by open walls in self-organizing systems
plays any relevant role far away from the wall, i.e.  whether the
boundaries induce long-range effects that might eventually alter the
bulk dynamics, affecting (bulk) universal critical properties.
Actually, if this was the case, then the understanding of SOC in terms
of standard non-equilibrium (bulk) phase transitions into absorbing
states \cite{FES} would be in jeopardize.

It is well known that, owing to the presence of diverging
correlations, walls induce non-trivial effects in critical phenomena,
i.e {\it surface critical phenomena} \cite{surface}. For instance, in
systems with a single absorbing state in the DP class, spreading
exponents \cite{AS} are known to differ from their bulk counterparts
if initial seeds of activity are localized in the neighborhood of a
wall \cite{DPwall}.  However, the universality class of the bulk
transition remains unaltered.

As we will show, in systems with a non-trivial background field (i.e.
systems with many absorbing states) walls can induce long-range
modifications of this field deep into the bulk (similar situations
have been addressed in the context of directed/anisotropic sandpiles
\cite{directed}), opening the possibility for relevant changes in the
bulk dynamics to occur.

In what follows, we analyze the effect of walls (both absorbing and
reflecting) in both Eq.(\ref{PCP}) and Eq.(\ref{CDP}) in {\it one
  dimensional} systems. First, we perform mean-field analyses and
second we study the full problem employing computer simulations.
Finally, we discuss the previous issues using the new insight.
 
\section{Mean Field results}
In mean field approximation the noise can be neglected (i.e.
$\sigma=0$), but in order to explore spatial structures we keep the
Laplacian terms. Absorbing boundaries are implemented as
\begin{equation}
  \rho(0,t) = 0, ~~~ \Phi(0,t) = 0 
  \label{abs}
\end{equation}
while reflecting ones correspond to
\begin{equation}
 \nabla (\rho(0,t))=0, ~~~ \nabla(\Phi(0,t))=0
  \label{ref}
\end{equation}
(i.e. Dirichlet and Neumann conditions, respectively), where $\Phi$
stands for the background field: $\Psi$ or $E$.
All the forthcoming discussions assume implicitly that a well-defined
stationary state exists. For this reason, we approach the critical
point from the active phase (to avoid getting trapped into absorbing
configurations, which prevents the system from relaxing to a true
self-averaging steady state).

\vspace{0.75cm} {\bf DP class with many AS:} The analysis of
Eq.(\ref{PCP}) with $\sigma=0$ becomes trivial if one subtracts the
second equation from the first one (multiplied by $D$) and assumes
stationarity.  In this way, the discretized Laplacian terms (either at
the wall or away from it) cancel out and one obtains a
site-independent equation, leading to spatially homogeneous solutions
for either absorbing and reflecting boundaries.  

Note that we have analyzed Eq.(\ref{PCP}), which includes explicitly a
$\nabla^2 \rho(x,t)$ term in the equation for the background field.
Such a term, present at the coarse-grained Langevin theory
\cite{many}, is neglected in many studies interested only in critical
properties. Actually, it can be argued to be irrelevant in the
renormalization group sense; still, as we have seen, it is important
to study spatial properties.
\begin{figure}
  \includegraphics[height=.25\textheight]{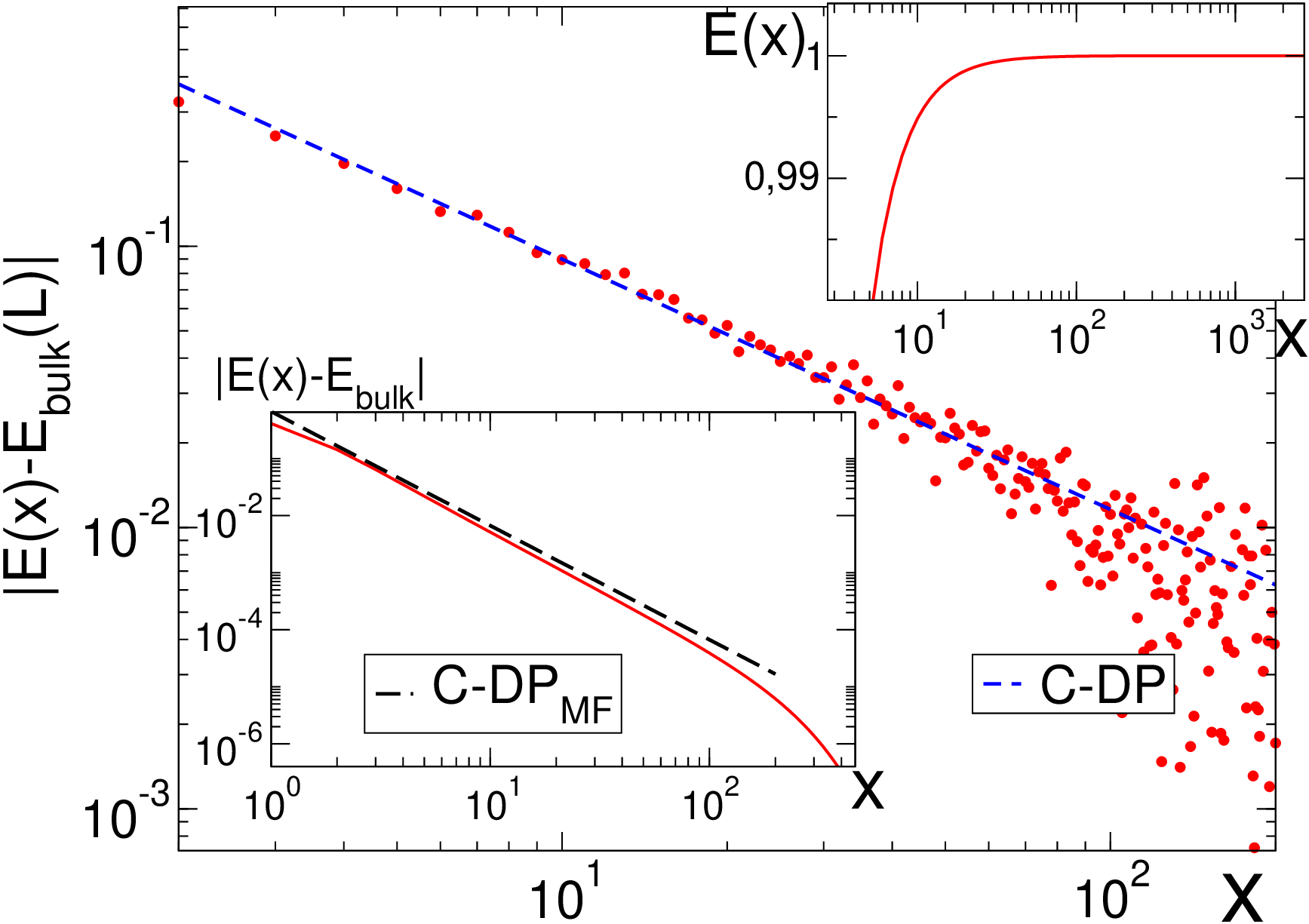}
  \includegraphics[height=.25\textheight]{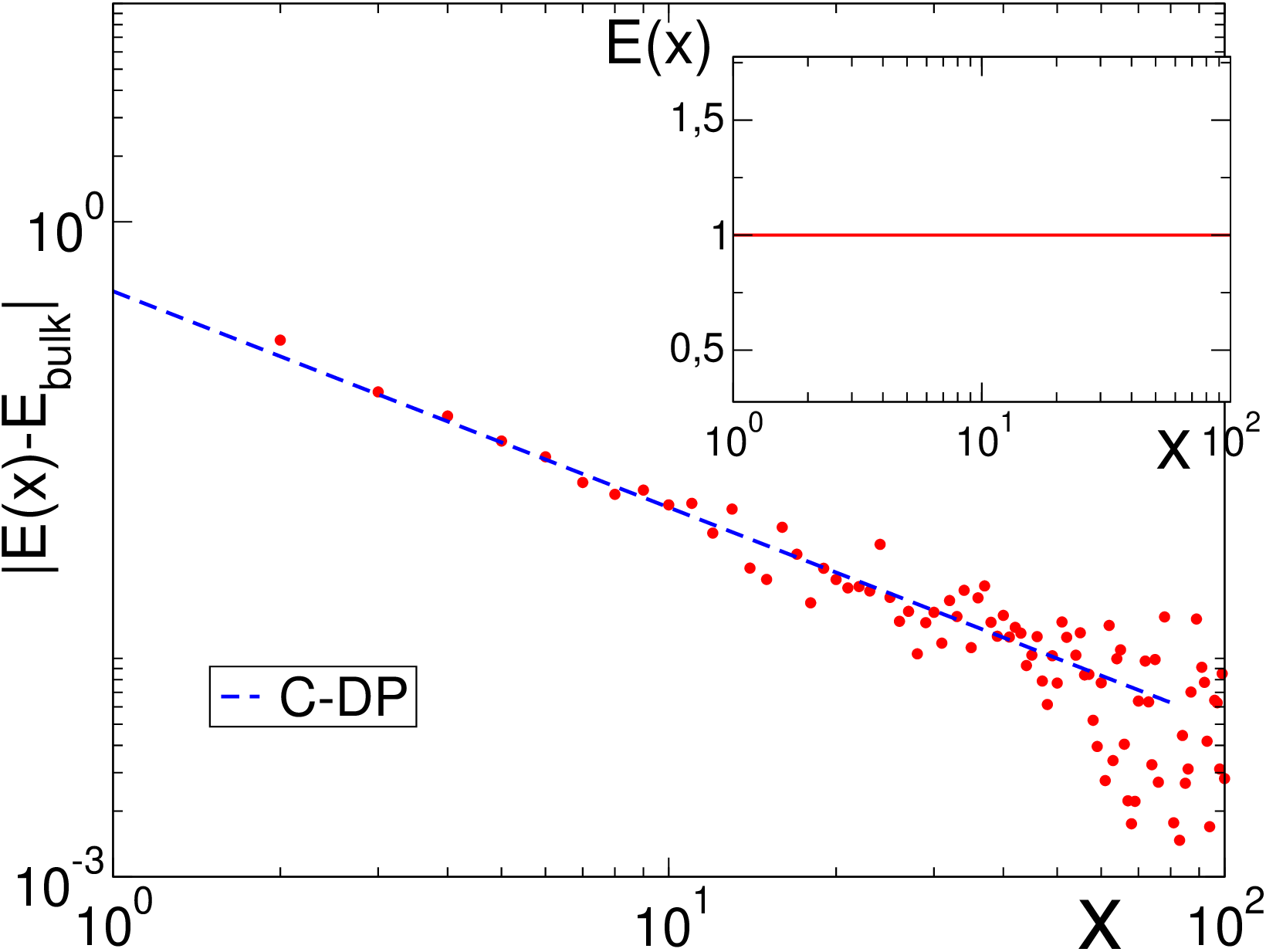}
  \caption{Background field for the C-DP equations, Eq.(\ref{CDP}),
    with an absorbing (left) and a reflecting boundary (right),
    respectively. The main plots correspond to the numerical
    integration of the full set of equations; in both cases a
    power-law decay is observed for $|E(x,t)-E_{bulk}|$. The two
    smaller insets show mean field results: a structured background in
    the absorbing case (left) and a flat one for reflecting walls
    (right).  The large inset in the left figure, illustrates the
    power-law decay of $|E(x,t)-E_{bulk}|$ in the mean field solution
    for absorbing walls at a finite time.}
\end{figure}

\vspace{0.75cm} {\bf C-DP:} Imposing (Neumann) reflecting conditions,
and proceeding as above (i.e. subtracting one equation from the other
to get rid of the Laplacians) it is straightforward to see that the
only possible steady state is a flat one, as in the DP case.

On the other hand, considering absorbing boundaries, the problem
becomes more subtle. Integrating the background equation in space, the
total amount of energy decreases as $-\nabla \rho(0,t) < 0$ , which
implies that there is a ``leakage of energy'' at the open boundary
and, hence, the only true steady state is the trivial one
$\rho(x)=E(x)=0$.

However, for long but finite times and starting from a flat initial
state, there is a non-trivial profile matching the two boundaries:
$E(0,t)=0$ and $E(\infty,t)=1=E_{bulk}$, where $E_{bulk}$ corresponds
to the initial condition. A simple calculation allows to derive the
shape of this landscape. We look for a solution using the following
ansatz: $E(x,t)= E_{bulk} - x^{-\alpha} \mathcal{F}(x^2/t) $, where
$\alpha$ is some exponent and $\mathcal{F}(x^2/t)$ an unknown function
of the scaling variable $x^2/t$. As the linear coefficient for the
activity equation is linear in $E$, we can try a solution of the form
$\rho(x,t)=\rho_{bulk} - x^{-\alpha} \mathcal{G}(x^2/t)$.  Plugging
this into the equation for the activity at the bulk critical point,
$a~=~-w~E_{bulk}$ (for which $\rho_{bulk}=0$) and equating the lowest
orders, it follows that $\alpha=2$ and that $\mathcal{G}(x^2/t)$ needs
to be of the form $ \frac{x^2}{t} {\mathcal{F}}(x^2/t) $ for
a solution to exist.  For asymptotically large times the leading order
of such a calculation gives rather straightforwardly
\begin{eqnarray}
 \rho(x,t) & =& t^{-1} {\mathcal{F}}(x^2/t) \\
 E(x,t) - E_{bulk} & =&  x^{-2} \mathcal{F}(x^2/t) 
 \label{scaling} \end{eqnarray}
where $\mathcal{F} $ is a degenerate hypergeometric  function.
In conclusion: there is a non-stationary structured background for any
finite time.  It consists of a power-law with exponent $\alpha=2$
converging asymptotically in the bulk to the closed-boundaries initial
value $E_{bulk}$, multiplied by a scaling function of $x^2/t$
(see Eq.(\ref{scaling})).

The exponent $\alpha=2$ can also be derived using na\"ive
power-counting arguments. As said above, the background field
contributes linearly to the coefficient of the linear term in the
activity equation, it behaves as the distance to the critical point
and, therefore, it scales with the inverse of the correlation length
critical exponent, $\nu$, which in mean-field approximation is
$\nu=1/2$ \cite{AS}, entailing $\alpha=2$.

Numerical integration of Eq.(\ref{CDP}) with $\sigma=0$ (i.e. in its
noiseless version) confirms these conclusions (see the insets of
Fig.1a and Fig. 1b, for absorbing and reflecting conditions,
respectively).

\section{Beyond mean field} In order to go beyond mean field
approximation, we switch-on back the noise term both in Eq.(\ref{PCP})
and Eq.(\ref{CDP}). Analytical solutions do not exist anymore and we
need to resort to numerical simulations in one-dimensional discretized
lattices.  Absorbing and reflecting boundaries correspond,
respectively to: $f(-1)=f(0)=0$ and $f(-1)=f(1)$ (where $f$ stand for
either $\rho, \Psi$, or $E$).  Using this, the discretized Laplacian
operator, $\nabla^2 f(x)= f(x+1)+f(x-1)- 2 f(x)$ can be replaced, at
the wall, by $\nabla^2 = \nabla - I$ (where $I$ is the identity) for
absorbing walls, and $\nabla^2 = 2 \nabla$ for reflecting ones.
Integration is performed by using the scheme proposed recently by
Dornic et al \cite{Dornic}, which allows to integrate square-root
noise stochastic equations in an efficient way.  We consider systems
of different sizes $L$ (from $2^{8}$ to $2^{12}$).

For each of them we determine numerically the critical point (its
location is slightly affected by finite size effects).  Contrarily to
the mean field case, now, the activity can reach the absorbing state
$\rho(x)=0$ in finite time and, hence, a steady state for the
background field exists for both absorbing or reflecting boundaries,
as we will illustrate. In numerics, a steady state is achieved by
perturbing with some small amount of activity the background field,
leaving the system relax to a new absorbing configuration, and
iterating this procedure as much as needed.

\vspace{0.75cm} {\bf DP class with many AS:} Numerical integration of
Eq.(\ref{PCP}) shows results qualitatively similar to those obtained
in the mean-field approach. For both absorbing and reflecting walls
there is an extremely fast convergence (it involves only a few sites)
from $E(0)=0$ to $E=E_{bulk}$ (results not shown).
\begin{figure}
  \includegraphics[height=.25\textheight]{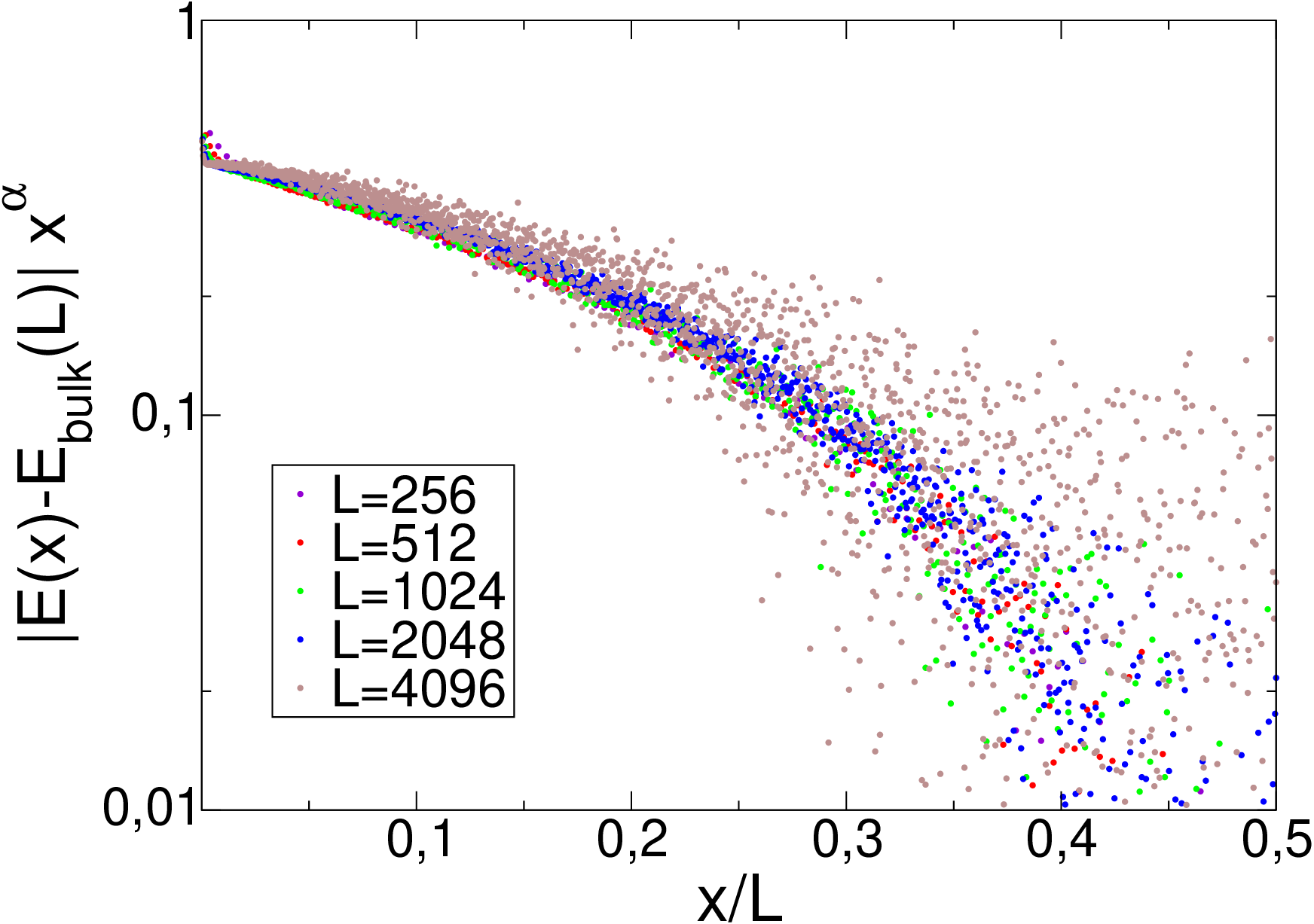}
 \includegraphics[height=.25\textheight]{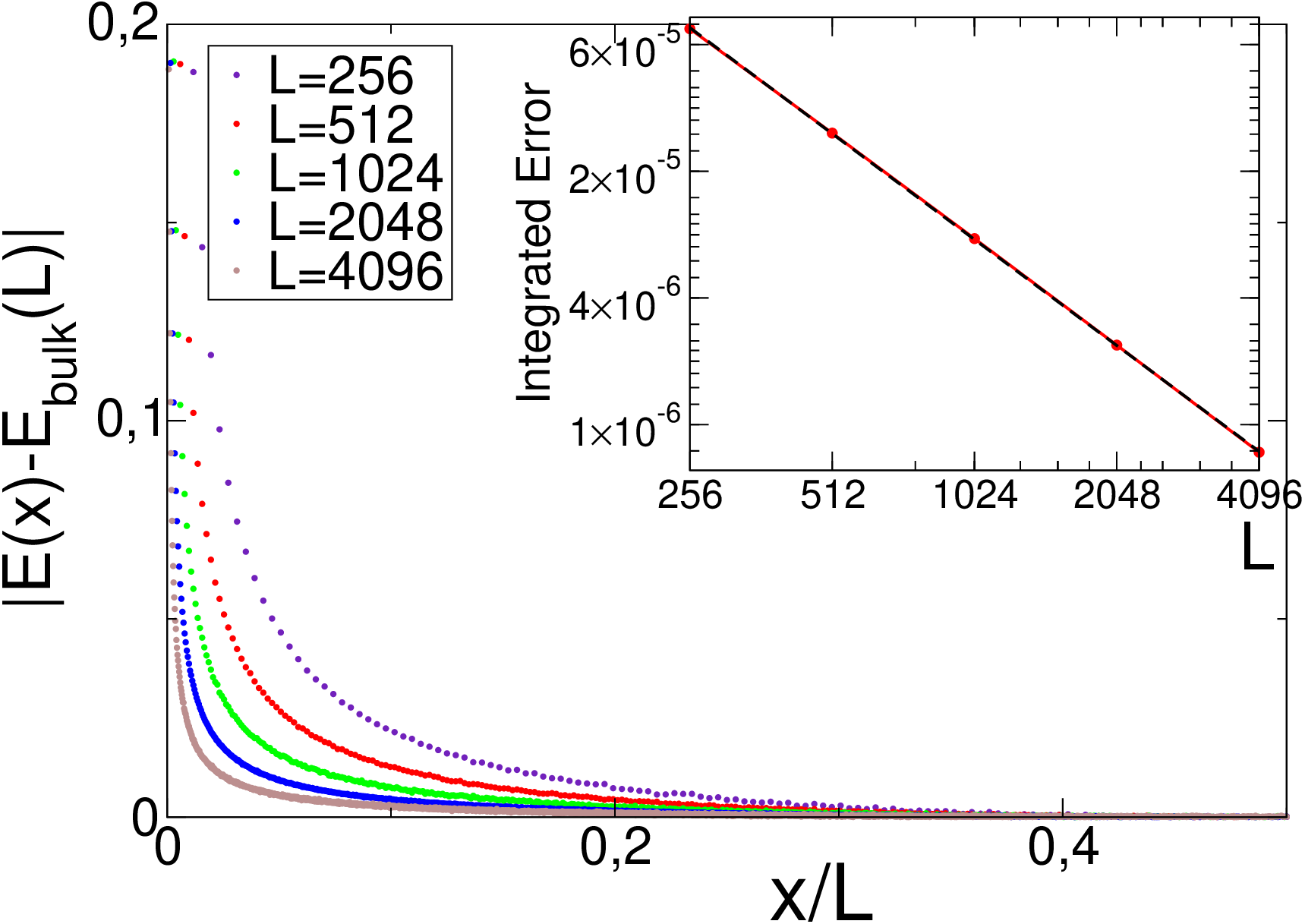}
 \caption{Left: Collapsed background fields, $|E(x)-E_{bulk}|
   x^{\alpha}$, as a function of $x/L$ for a system into the C-DP universality class,
   with an absorbing boundary, at criticality, and for
   different sizes.  Using $\alpha= 0.80(5)$ all curves
   collapse into a unique one.  Right: Stationary background field for
   the same C-DP system at criticality, for different
   values of $L$. The inset shows a plot of the area below the
   different curves, normalized by system size, as a function of $L$. }
\end{figure}

\vspace{0.75cm} {\bf C-DP:} Figure 1(main plots) shows results for the
stationary backgrounds coming out of a numerical integration of
Eq.(\ref{CDP}) for absorbing and reflecting walls respectively
($L=4096$). Both log-log plots show a power-law convergence to the
bulk value. In the absorbing case there is an under-density nearby the
wall, while in the reflecting case there is an over-density.

One might wonder why arguments as those presented in the previous
section to discard the possibility of structured backgrounds in the
reflecting case fail here. More specifically: if the background field
is to be stationary on average, then $\nabla^2 \rho$ cannot have any
structure on average (see Eq.(\ref{CDP})). From this simple relation,
in mean-field approximation we concluded that there cannot be a
non-trivial structure neither for $\rho$ nor for $E(x)$.  Instead, in
the presence of fluctuations, even if $\rho(x)$ does not have any
non-trivial spatial structure (i.e. even if $\nabla^2 \rho$ vanishes
on average), $\rho^2(x)$, $\sqrt{\rho(x)})$ and $E(x)$ can have one
and actually, we have verified in numerical simulations that they do
have one. In other words; even if the first moment of $\rho$ is
structureless, the second one and the average of its square-root take
different values at different sites, having therefore a non-trivial
structure.  This can happen only for fluctuating variables implying
that the non-trivial structure in this case is a {\it noise-induced
  one}.

Fig.2a illustrates that, in the absorbing case, the backgrounds for
different sizes collapse into a unique scaling curve by assuming
$|E_{bulk}(L)-E(x)| \sim x^{-\alpha} \mathcal{F}(x/L)$ with
$\alpha=0.80(5)$.  Using scaling arguments as the mean-field ones
above, $\alpha$ should coincide with $1/\nu$.  $\nu$ is known from
previous work (and our own direct measurements) to be $\nu \approx
1.33(5)$ \cite{Lubeck}, implying $\alpha \approx 0.75(5)$, compatible
with our previous estimation.  Note that this solution coincides
qualitatively with Eq.(\ref{scaling}), but the exponent $\alpha$ takes
the value $1/\nu =2$ in mean-field and its renormalized value $1/\nu
\approx 0.75$ here.

In Fig.2b we plot the deviation of the background field from its
asymptotic bulk value, $|E(x)-E_{bulk}|$, as a function of $x/L$.  For
all values of $L$, the decay exhibits a fat tail. However, as shown in
the inset of Fig.2b, the global deviation from the bulk value, defined
as the spatial integral of $|E_{bulk}-E(x)|$ divided by $L$, decays
exponentially fast to $0$ as $L$ is increased, i.e.  $\int dx
|E_{bulk}-E(x)|$ is sub-extensive.  In other words, {\it the global
  effect of an absorbing boundary on the bulk dynamics becomes
  arbitrarily small for sufficiently large system sizes; this
  guarantees the existence of a well defined bulk in the thermodynamic
  limit}.  In other words, the bulk behavior in the thermodynamic
limit can be systematically approached by considering a sequence of
finite-size open-boundaries systems with larger and larger sizes.
Identical results are obtained in the reflecting case (not shown).

\section{Discussion and Conclusions}

The summary of the previous findings is as follows.  Models in the DP
class have a mostly non-structured background field at criticality
both for absorbing and reflecting walls.
This result is obtained in a mean field calculation as well as in
numerical simulations. Walls in the DP class affect boundary/surface
properties, but not bulk ones.

On the other hand, systems with a conservation-law (in the C-DP class)
exhibit, for both absorbing and reflecting walls, a power-law
convergence to their corresponding bulk value of the background field.
This stems from the purely diffusive and conserved nature of the
corresponding background-field equation.

The absorbing case is qualitatively well described by our mean-field
(noiseless) approach (which leads to a local under-density nearby the
wall, converging as a power law to the bulk value), even if with an exponent
different from the mean-field one.

Instead, in the reflecting case, the mean-field calculation does not
lead to any non-trivial structure. This is so because the derivative
of the background field at the boundary is fixed to $0$ and this leads
ineluctably to a flat background.  Instead, in the presence of
fluctuations, locally non-vanishing derivatives of the field appear at
the boundary; these are then amplified, a local over-density of the
background field is generated, and (in the long time limit) a
power-law decaying stationary background structure sets in.
Therefore, once fluctuations are switched on, a {\it noise-induced
  non-trivial background structure emerges}.

For both absorbing and reflecting backgrounds the spatial convergence
to the corresponding bulk value is described by a power law with an
exponent $\alpha$ equal to the inverse of the (bulk) correlation
length exponent.

Finally, let us emphasize that a well defined bulk exits in all cases
owing to the fact that the global deviation of the averaged background
field with respect to its bulk value converges exponentially fast to
zero in the large system size limit (as illustrated in Fig.2b). It is
important to underline that this does not imply that {\it boundary
  critical exponents} cannot be affected by the presence of walls, and
actually they typically are \cite{Jabo2}.  For instance, it is well
known that walls change the surviving probability for the propagation
of activity from a localized seed nearby the boundary, affecting
spreading exponents. Instead, bulk properties are not affected in any
case, despite of the power-law convergence of the background reported
above. The global effect of the wall on bulk properties can be made as
small as wanted by enlarging the system size, implying the existence
of a well defined bulk in the thermodynamic limit.  This provides
further conceptual support for the understanding of self-organizing
sandpiles (with open boundaries) from the perspective of standard
phase bulk transitions in systems with many absorbing states and a
conservation law \cite{FES}.  \vspace{-0.25cm}

\begin{theacknowledgments}
  We acknowledge useful and motivating discussions with D. Dhar, P.K.
  Mohanty, and P.L. Garrido.  Support from the Spanish MEyC-FEDER,
  project FIS2005-00791, and from Junta de Andaluc{\'\i}a as group
  FQM-165 is acknowledged.
\end{theacknowledgments}

\end{document}